\documentclass[amsmath,amstex,eqsecnum,preprint,pre,aps]{revtex4}
\usepackage{amsmath}
\usepackage{graphicx}
\usepackage{dcolumn}
\usepackage{bm}
\usepackage{float}

\begin{document}
\title[Overdamped van Hove function of atomic liquids]{Overdamped van Hove function of atomic liquids}

\author{ Leticia L\'opez-Flores$^1$,   Laura L. Yeomans-Reyna$^{2}$, Mart\'in
Ch\'avez-P\'aez$^3$, and Magdaleno Medina-Noyola$^3$}

\address{$^1$Facultad de Ciencias Fisico-Matem\'aticas,
Benem\'{e}rita Universidad Aut\'{o}noma de Puebla, C.P.72000, Puebla, Pue., M\'{e}xico}

\address{$^{2}$ Departamento de F\'{\i}sica, Universidad de Sonora, Boulevard Luis \\
Encinas y Rosales, 83000, Hermosillo, Sonora, M\'{e}xico.}

\address{$^{3}$Instituto de F\'{\i}sica {\sl ``Manuel Sandoval Vallarta"},
Universidad Aut\'{o}noma de San Luis Potos\'{\i}, \'{A}lvaro
Obreg\'{o}n 64, 78000 San Luis Potos\'{\i}, SLP, M\'{e}xico}

\date{\today}

\begin{abstract}

Using the generalized Langevin equation formalism and the process of contraction of the description we derive a  general memory function equation for the thermal fluctuations of the local density of a simple atomic liquid. From the analysis of the long-time limit of this equation, a striking equivalence is suggested between the long-time dynamics of the atomic liquid and the dynamics of the corresponding \emph{Brownian} liquid. This dynamic equivalence is confirmed here by comparing molecular and Brownian dynamics simulations of the self-intermediate scattering function and the long-time self-diffusion coefficient for the hard-sphere liquid.

\end{abstract}


%\submitto{\JPCM}
\maketitle

\section{Introduction.}\label{sectionI}

In many respects, the properties of colloidal fluids resemble almost perfectly those of the correspondent atomic liquid \cite{pusey0,nagele0,deschepperpusey1, deschepperpusey2}.
It is well known that the equilibrium phase diagram, and in general all the equilibrium thermodynamic properties, of a specific model system (say a Lennard-Jones liquid) will be independent of the microscopic (either molecular or Brownian) dynamics that govern the motion of the $N$ interacting particles that constitute the system. This implies that these equilibrium properties can be generated using either molecular or Brownian dynamics simulations \cite{tildesley}. Furthermore, although time-dependent properties
are expected in general to depend on the specific microscopic
dynamics, some features associated with the
long-time dynamic behavior of the system also seem to be rather insensitive
to the microscopic short-time dynamics. This appears to be particularly true regarding the rather complex dynamic behavior of these systems as they approach the glass transition \cite{lowenhansenroux,szamelflenner,puertasaging}. Determining the range of validity of this analogy continues to be a relevant topic in the study of the dynamics of liquids.

From the theoretical side one would like to unify  colloidal
and atomic liquids in a common theoretical description of the relaxation dynamics of the local density fluctuations,  which explicitly exhibits the origin of the
similarities and differences in their macroscopic dynamics. One possible general framework for such theoretical analysis is the concept of the generalized Langevin equation (GLE) \cite{delrio,faraday}. This equation describes the dynamics of the thermal fluctuations $\delta a_{i}(t)\ (\equiv a_{i}(t)-a^{eq}_{i})$ of the instantaneous value of the macroscopic variables $ a_{i}(t)$ ($i=1,2,...,\nu$), around its equilibrium value $a^{eq}_{i}$, and has the structure of the most general linear stochastic equation with additive noise for the vector $\delta \mathbf{a}(t)=\left[\delta a_{1}(t),\delta a_{2}(t),...,\delta a_{\nu }(t)\right]^{\dagger} $ (with the dagger meaning transpose). The GLE equation has been widely used in the description of thermal fluctuation phenomena in simple liquid systems, and Boon and Yip's textbook \cite{boonyip} contains a detailed account of its early use to describe the dynamics of simple liquids. Although this stochastic equation is conventionally associated with the Mori-Zwanzig projection operator formalism \cite{zwanzig,mori}, in reality its structure is not a consequence of the hamiltonian basis of Mori-Zwanzig's derivation; instead, it is essentially equivalent to the mathematical condition of stationarity \cite{delrio}.

Thus, in Ref. \cite{scgle0} the GLE formalism, understood in the latter manner,  was employed to derive the most general diffusion equation of a model Brownian liquid (i.e., an idealized monodisperse colloidal suspension in the absence of hydrodynamic interactions)  formed by $N$ spherical Brownian particles interacting between them through direct (i.e., conservative) forces, but in the absence of hydrodynamic interactions. The resulting general memory function expression for the intermediate scattering function (ISF) $F(k,t)$ and for its self component $F_S(k,t)$, were later employed in the construction of the self-consistent generalized Langevin equation (SCGLE) theory of colloid dynamics \cite{scgle1,scgle2}, eventually applied to the description of dynamic arrest phenomena \cite{rmf,todos1,todos2} and more recently \cite{noneqscgle0,noneqscgle1} to the construction of a first-principles theory of equilibration and aging of colloidal glass-forming liquids.

With the aim of investigating the relationship between the dynamics of atomic and Brownian liquids, here we start the extension of these theoretical developments to describe the macroscopic dynamics of both kinds of systems within the same theoretical formalism. With this general intention in mind, in the present paper we discuss the application of the generalized Langevin equation formalism above, to the derivation of general memory-function expressions for the (collective and self) intermediate scattering functions of an atomic liquid. These expressions should in principle be capable of describing the crossover behavior of these properties between their ballistic short time limit and their diffusive long-time behavior. Although in practice we do not use these expressions here to numerically evaluate these functions in the short- or intermediate time-regime $t\approx \tau_0$ (where $ \tau_0$ is the mean free time), we find that in their long-time limit, $t \gg \tau_0$, these expressions for  $F(k,t)$ and $F_S(k,t)$ become essentially identical to the corresponding expressions for a colloidal fluid, strongly suggesting a well defined long-time dynamic correspondence between atomic and colloidal liquids.

The strategy that we shall employ to derive the memory function equations for the intermediate scattering functions of our model atomic liquid will actually rely very heavily on the referred previous derivation \cite{scgle0} of the time-evolution equations for $F(k,t)$ and $F_S(k,t)$ of the corresponding idealized  Brownian fluid. The rationale  for this is the rather simple observation that the essential difference between an atomic liquid  and its idealized Brownian counterpart (a colloidal liquid in the absence of hydrodynamic interactions) is the presence, in the microscopic equations of motion of the latter, of the  friction force $-\zeta ^{(s)}{\bf v}_{i}(t)$  due to the supporting solvent and the corresponding fluctuating force ${\bf f}^{(s)}(t)$. Thus, we first review the derivation of Ref. \cite{scgle0}, with the aim of keeping track of the effects of these friction terms. This aspect of the present work is developed in section \ref{sectionII}. At the end of the section, we simply take the $\zeta ^{(s)}\to 0$ limit of the end result of the referred derivation, to obtain the corresponding  time-evolution equations for $F(k,t)$ and $F_S(k,t)$ of our atomic liquid (namely, Eqs. (\ref{fdkz0}) and (\ref{fsdkz0})).

The next task of this work is to analyze the long-time limit of these results for $F(k,t)$ and $F_S(k,t)$. In Ref.  \cite{scgle0}, dealing with Brownian systems, this limit was referred to as the ``overdamped" limit, corresponding to times $t$ much longer than the relaxation time  $\tau ^{(s)} \equiv M/\zeta ^{(s)}$ of the velocity autocorrelation function. This relaxation results from the damping of the particle's momentum due to the  friction force $-\zeta ^{(s)}{\bf v}_{i}(t)$. Thus, in that case  $\tau ^{(s)}$ sets the crossover timescale from the early initial regime $t << \tau^{(s)}$, where the inertial effects are still important, to the long-time regime  $t>> \tau ^{(s)}$, where the motion of the suspended particles is purely diffusive, and described by the short-time self-diffusion coefficient $D^{(s)}=k_BT/\zeta ^{(s)}$. In contrast, in atomic liquids an analogous timescale is apparently absent, since there is not any material solvent exerting damping friction forces. In spite of that, in Section \ref{sectionIII}, we analyze the long-time limit of the time-evolution equations for $F(k,t)$ and $F_S(k,t)$ of the atomic liquid derived in Section \ref{sectionII}. We find that in this limit, these equations happen to adopt the same structure as the corresponding equations for Brownian systems in their overdamped limit. As a result of this analysis, we conclude that the parameter playing the role of the short-time self-diffusion coefficient $D^{(s)}$ is now the self-diffusion coefficient $D^{0}$ determined by kinetic theory.

This formal dynamic correspondence has important physical consequences, expressed in terms of well defined scaling properties of the dynamics of two fluid systems which only differ in the microscopic laws  that govern the motion of the constituent particles (either molecular or Brownian dynamics). The most relevant of such consequences are briefly discussed in the final section (Section \ref{sectionV}) of this paper.

\section{Atomic fluid as a frictionless Brownian liquid.} \label{sectionII}

Let us start by reviewing the derivation in Ref. \cite{scgle0} of the time-evolution equations of $F(k,t)$ and $F_S(k,t)$ of an idealized monodisperse colloidal suspension in the absence of hydrodynamic interactions, formed by $N$ spherical particles in a
volume $V$, whose microscopic dynamics is described by the $N$-particle Langevin equations \cite
{3,4,5}
\begin{equation}
M{\frac{d{\bf v}_{i}(t)}{dt}}\equiv -\zeta ^{(s)}{\bf v}_{i}(t)+{\bf f}^{(s)}
_{i}(t)+\sum_{j\neq i}{\bf F}_{ij}(t),\quad (i=1,2,\ldots ,N).  \label{eq1}
\end{equation}
In these equations, $M$ is the mass and ${\bf v}_{i}(t)$ the
velocity of the $i$th particle, and $\zeta^{(s)}$ is its friction coefficient
in the absence of interactions. Also, ${\bf f}^{(s)}_{i}(t)$ is a random force,
modeled as a Gaussian white noise of zero mean, and variance given by $%
\langle {\bf f}^{(s)}_{i}(t){\bf f}^{(s)}_{j}(0)\rangle =k_{B}T\zeta^{(s)}2\delta
(t)\delta _{ij}\stackrel{\leftrightarrow }{{\bf I}}(i,j=1,2,\ldots ,N;%
\stackrel{\leftrightarrow }{{\bf I}}\mbox{being the }3\times 3%
\mbox{
unit tensor})$. The direct interactions between the particles are
represented by the sum of the pairwise forces ${\bf F}_{ij}$ that the $j$th
particle exerts on particle $i$, i.e., ${\bf F}_{ij}$ is obtained from the
pair potential $u(|{\bf r}_{i}-{\bf r}_{j}|)$.

Our goal is to derive the macroscopic time-evolution equations for the ISFs $F(k,t)$ and $F_{S}(k,t)$, starting from this microscopic level of description. Some of the most important features of such general time evolution equations for $F(k,t)$ and $F_{S}(k,t)$   can be written, however, right at the outset, since they derive from
the general selection rules \cite{delrio} originating from the stationarity
condition and from other symmetry properties of the macroscopic variables
whose dynamics couple to the dynamics of the local particle concentration.
This was the approach adopted in Ref. \cite{scgle0}, which derived the most general
time-evolution equation for the fluctuations of the local concentration $n({\bf r},t)$ \ of colloidal particles, consistent with the selection rules referred to above. The specific information of the microscopic
dynamics, was then employed in the approximate or partial determination of
those elements of the time-evolution equation that such selection rules
left undetermined. This section briefly summarizes the main steps of such derivation.

At each step of the following derivation, however, we urge the reader to keep track of the particular case in which the friction term $-\zeta ^{(s)}{\bf v}_{i}(t)$ and its corresponding fluctuating force ${\bf f}^{(s)}
_{i}(t)$ are absent, and to recognize that an
\emph{atomic} liquid can be viewed as the present Brownian liquid in the
limit of an infinitely tenuous solvent, such that the Stokes
friction coefficient $\zeta ^{(s)}$ vanishes. Thus,  we shall take the limit  $\zeta ^{(s)} \to 0$ in the general memory-function expressions for  $F(k, t)$ and $F_S(k,t)$ derived in this section. In such limit one is left
only with the particles in the vacuum, and
these equations  for  $F(k, t)$ and $F_S(k,t)$ will then become the exact memory function expressions for the ISFs of an \emph{atomic} liquid.

Thus, let us first recall that the basis of the GLE formalism are the general mathematical conditions stated by the theorem of stationarity \cite{delrio}. This theorem states that the equation describing the dynamics of the thermal fluctuations $\delta a_{i}(t)\ (\equiv a_{i}(t)-a^{eq}_{i})$ of the instantaneous value of the macroscopic variables $ a_{i}(t)$ ($i=1,2,...,\nu$) around its equilibrium value $a^{eq}_{i}$ must have the structure of the most general linear stochastic equation with additive noise for the vector $\delta \mathbf{a}(t)=\left[\delta a_{1}(t),\delta a_{2}(t),...,\delta a_{\nu }(t)\right]^\dagger $, namely,
\begin{equation}
\frac{d\delta \mathbf{a}(t)}{dt}=-\omega \chi ^{-1}\delta \mathbf{a}(t)-\int%
\limits_{0}^{t}L(t-t^{\prime })\chi ^{-1}\delta \mathbf{a}(t^{\prime })dt^{\prime }+%
\mathbf{f}(t).
\label{gle0}
\end{equation}
In this equation $\chi $ is the matrix of static correlations,  $\chi
_{ij}\equiv \left\langle \delta  a_{i}(0) \delta  a_{j}^{\ast }(0)\right\rangle $, $\omega $
is an anti-Hermitian matrix ($\omega _{ij}=-\omega _{ji}^{\ast }$), and the matrix $L(t)$
is determined by the fluctuation-dissipation relation $L_{ij}(t)=\left\langle
f_{i}(t)f_{j}^*(0)\right\rangle $, where $f_{i}(t)$ is the $i$th component of the vector of random forces $\mathbf{f}(t)$.

For the present purpose, we choose the components of the state vector $\delta {\bf a}(t)$ as
\begin{equation}
\delta {\bf a}(t)\equiv \left[ \delta n({\bf k},t),\delta j({\bf k},t),\delta
\sigma _{K}({\bf k},t),\delta \sigma _{U}({\bf k},t)\right]^{\dagger},
\end{equation}
with the following definitions. First, $a_{1}(t)$ is the Fourier transform $\delta n({\bf k},t)$ of the fluctuations $\delta n({\bf r},t)\equiv n({\bf r},t)-n$ of the local concentration $n({\bf r},t)$ around
its bulk value $n$. The microscopic definition  of $\delta n({\bf k},t)$ (for ${\bf k}=0$) is
\begin{equation}
\delta n({\bf k},t)=\frac{1}{\sqrt{N}}\sum_{i=1}^{N}e^{i{\bf k\cdot r}_{i}(t)}, \label{defnumber}
\end{equation}
where ${\bf r}_{i}(t)$ is the position of the $i$th colloidal particle at
time $t$. Normalized in this manner $\delta n({\bf k},t)$ is such that its
static correlation is $\chi _{nn}(k)\equiv \left\langle \delta n(%
{\bf k},0)\delta n(-{\bf k},0)\right\rangle =S(k)$, where $S(k)$ is the
static structure factor of the bulk suspension.

Taking the time-derivative of $\delta n({\bf k},t)$ we have the continuity equation,
\begin{equation}
\frac{\partial \delta n({\bf k},t)}{\partial t}=ik\delta j_{l}({\bf k},t), \label{continuity}
\end{equation}
where $\delta j_{l}({\bf k},t)\equiv j_{l}({\bf k},t)=\widehat{{\bf k}}{\bf %
\cdot j}({\bf k},t)$ is the component of the current ${\bf j}({\bf k},t)$ in
the direction $\widehat{{\bf k}}$ of the vector ${\bf k}$, i.e.,
\begin{equation}
j_{l}({\bf k},t)=\frac{1}{\sqrt{N}}\sum_{i=1}^{N}\widehat{{\bf k}}{\bf \cdot
v}_{i}(t)e^{i{\bf k\cdot r}_{i}(t)} \label{defcurrent}
\end{equation}
with ${\bf v}_{i}(t)=d{\bf r}_{i}(t)/dt$. Thus,  $a_2
(t)\equiv \delta j_{l}({\bf k},t)$, whose
static correlation matrix  is
\begin{equation}
\chi _{jj}=k_{B}T/M. \label{chijj}
\end{equation}

If we take the time-derivative of the current in Eq. (\ref{defcurrent}), and employ the $N$-particle
Langevin equation, Eq. (\ref{eq1}), we are led to the following result
\begin{equation}
\frac{\partial \delta j_{l}({\bf k},t)}{\partial t}=-\frac{\zeta^{(s)}}{M}%
\delta j_{l}({\bf k},t)+\frac{f^{(s)}({\bf k},t)}{M}+\frac{1}{\sqrt{N}}%
\sum_{i=1}^{N}\widehat{{\bf k}}{\bf \cdot }\frac{{\bf F}_{i}(t)}{M}e^{i{\bf %
k\cdot r}_{i}(t)} +\frac{ik}{\sqrt{N}}\sum_{i=1}^{N}\left[ \widehat{{\bf k}}%
{\bf \cdot v}_{i}(t)\right] ^{2}e^{i{\bf k\cdot r}_{i}(t)}
\end{equation}
where
\begin{equation}
f^{(s)}({\bf k},t)\equiv \frac{1}{\sqrt{N}}\sum_{i=1}^{N}\widehat{{\bf k}}{\bf
\cdot f}^{(s)}_{i}(t)e^{i{\bf k\cdot r}_{i}(t)},
\end{equation}
and ${\bf F}_{i}(t)\equiv \sum_{j\neq i}{\bf F}_{ij}(t)$. This equation can also be written as
\begin{equation}
\frac{\partial \delta j_{l}({\bf k},t)}{\partial t}=-\frac{\zeta^{(s)}}{M}%
\delta j_{l}({\bf k},t)+\frac{f^{(s)}({\bf k},t)}{M}+ik\delta \sigma ^{zz}({\bf k},t),
\label{djdt0}
\end{equation}
with $\delta \sigma ^{zz}(k,t)$ being the
instantaneous fluctuation of the isotropic diagonal component of the stress
tensor
\begin{equation}
\sigma ^{_{\alpha \beta }}({\bf k},t)\equiv \frac{1}{\sqrt{N}}
\sum_{i=1}^{N}\left\{ v_{i}^{\alpha }v_{j}^{\beta }-\frac{1}{2M}\sum_{j\neq
i}\frac{r_{ij}^{\alpha }r_{ij}^{\beta }}{r_{ij}^{2}}P_{k}(r_{ij})\right\}
e^{i{\bf k\cdot r}_{i}(t)},
\end{equation}
where
\begin{equation}
P_{k}(r_{ij})\equiv r_{ij}\frac{du(r_{ij})}{dr_{ij}}\frac{e^{i{\bf k\cdot r}%
_{ij}(t)}-1}{{\bf k\cdot r}_{ij}(t)}.
\end{equation}
In these equations, ${\bf r}_{ij}\equiv {\bf r}_{i}-{\bf r}_{j}$, and $%
u(r_{ij})$ is the pair potential.

Let us now write $\delta \sigma ^{zz}({\bf k},t)$  as
\begin{equation}
\delta \sigma ^{zz}({\bf k},t)=\delta p({\bf k},t)+\delta \sigma _{K}({\bf k},t)+\delta \sigma _{U}({\bf k},t),
\end{equation}
with $\delta p(k,t)=[\chi _{jj}/S(k)]\delta n(k,t)$ being the Fourier transform of the local pressure fluctuations, and with  $\delta \sigma _{K}({\bf k},t)$ and  $\delta \sigma _{U}({\bf k},t)$ being the statically orthogonal kinetic and configurational components of [$\delta \sigma ^{zz}({\bf k},t)-\delta p({\bf k},t)]$,
defined as
\begin{equation}
\delta \sigma _{K}({\bf k},t)\equiv \frac{1}{\sqrt{N}} \sum_{i=1}^{N}(v_{i}^{z})^{2}e^{i{\bf k\cdot r}_{i}(t)}-\chi _{jj}\delta n({\bf k},t), \label{defsigmak}
\end{equation}
and
\begin{equation}
\delta \sigma _{U}({\bf k},t)\equiv -\frac{1}{2M\sqrt{N}}%
\sum_{i=1}^{N}\sum_{j\neq i}\frac{r_{ij}^{\alpha }r_{ij}^{\beta }}{r_{ij}^{2}
}P_{k}(r_{ij})e^{i{\bf k\cdot r}_{i}(t)}-\delta p({\bf k}
,t)+\chi _{jj}\delta n({\bf k},t). \label{defsigmau}
\end{equation}
This completes the microscopic definition of the components $\delta n({\bf k},t),\delta j({\bf k},t),\delta
\sigma _{K}({\bf k},t),$ and $\delta \sigma _{U}({\bf k},t)$ of the state vector ${\bf a}(t)$, which are then found in Eqs. (\ref{defnumber}), (\ref{defcurrent}), (\ref{defsigmak}), and (\ref{defsigmau}), respectively.

As a result, we finally rewrite the momentum conservation equation, Eq. (\ref{djdt0}), as
\begin{equation}
\frac{\partial \delta j_{l}({\bf k},t)}{\partial t}=-\frac{\zeta^{(s)}}{M}
\delta j_{l}({\bf k},t)+\frac{1}{M}f^{(s)}({\bf k},t)+ik\delta p({\bf k}
,t)+ik\delta \sigma _{K}({\bf k},t)+ik\delta \sigma _{U}({\bf k},t).
\label{djdt1}
\end{equation}
This equation, together with the continuity equation in Eq. (\ref{continuity}), couple the variables $\delta n({\bf k},t)$ and $\delta j({\bf k},t)$ with the variables $\delta
\sigma _{K}({\bf k},t)$ and $\delta \sigma _{U}({\bf k},t)$, whose time-evolution equation must now be determined, and the GLE formalism provides a natural manner to do
that. For this, one first performs a straightforward statistical thermodynamical
calculation of the matrix $\chi $ of static correlations $\chi _{ij}\equiv
\left\langle a_{i}(0)a_{j}^{\ast }(0)\right\rangle $, with the following result \cite{scgle0}
\begin{equation}
{\bf \chi }=\left[
\begin{array}{cccc}
\chi _{nn} & 0 & 0 & 0 \\
0 & \chi _{jj} & 0 & 0 \\
0 & 0 & \chi _{KK} & 0 \\
0 & 0 & 0 & \chi _{UU}
\end{array}
\right],
\end{equation}
with $\chi _{nn}=S(k)$ and $\chi _{jj}=k_{B}T/M$, and with $\chi
_{KK}$ and $\chi _{UU}$ given by
\begin{equation}
\chi _{KK}=2\chi _{jj}^{2}
\end{equation}
and
\begin{equation}
\chi _{UU}=\chi _{jj}^{2}\left[ 1+n\int d{\bf r}g(r)\frac{\partial ^{2}\beta
u(r)}{\partial z^{2}}\left( \frac{1-\cos (kz)}{k^{2}}\right) -\frac{1}{S(k)}%
\right].
\end{equation}

We then write up the generalized Langevin equation for our vector $\delta {\bf a}(t)$ in the format
of Eq. (\ref{gle0}). For this, we first
notice that all the variables, except $\delta a_{2}(t)=\delta j_{l}({\bf k},t)$, are
even functions under time-reversal. According to Onsager reciprocity
relations, and the general anti-hermiticity of $\omega $ and hermiticity of $
L(z)$  \cite{delrio}, we have that the only possibly non-zero elements of the
matrix $\omega $ and $L(z)$ are

\begin{equation}
\omega {\bf =}\left[
\begin{array}{cccc}
0 & \omega _{nj} & 0 & 0 \\
-\omega _{nj}^{\ast } & 0 & \omega _{jK} & \omega _{jU} \\
0 & -\omega _{jK}^{\ast } & 0 & 0 \\
0 & -\omega _{jU}^{\ast } & 0 & 0
\end{array}
\right]
\end{equation}

\begin{equation}
L(t)=\left[
\begin{array}{cccc}
L_{nn} & 0 & L_{nK} & L_{nU} \\
0 & L_{jj} & 0 & 0 \\
L_{nK}^{\ast } & 0 & L_{KK} & L_{KU} \\
L_{nU}^{\ast } & 0 & L_{KU}^{\ast } & L_{UU}
\end{array}
\right]
\end{equation}

The determination of the non-zero elements of $\omega $ and of some of the
non-zero elements of $L(t)$ is rather straightforward, since, from the exact
continuity equation,
\begin{equation}
\frac{\partial \delta n({\bf k},t)}{\partial t}=ik\delta j_{l}({\bf k},t)
\end{equation}
we immediately see that $\omega _{nj}=-ik\chi _{jj}$, and that $%
L_{nn}=L_{nK}=L_{nU}=0$. Similarly, from eq. (\ref{djdt1}) we can see that $\omega _{jK}\chi _{jU}^{-1}=\omega _{jU}\chi _{UU}^{-1}=-ik$
and $L_{jj}\chi _{jj}^{-1}=\zeta^{(s)}/M$. As a result, all the elements of
the ``frecuency'' matrix $\omega $ have been determined, and in fact, only
the kinetic coefficients $L_{KK}(k,z)$, $L_{KU}(k,z)=L_{UK}(k,z)$, and $%
L_{UU}(k,z)$ remain undetermined by general symmetry principles, or physical
principles such as mass or momentum conservation. Thus, the time-evolution
equations that complete the non-contracted description for the components of
the vector $\delta {\bf a}(t)$ are the mass and momentum conservation
equations, Eqs. (\ref{continuity}) and (\ref{djdt1}), along with the time-evolution equations for
$\delta \sigma _{K}({\bf k},t)$ and $\delta \sigma _{U}({\bf k},t)$, namely,
\begin{eqnarray}
\frac{\partial \delta \sigma _{K}({\bf k},t)}{\partial t} &=&ik\chi
_{KK}\chi _{jj}^{-1}\delta j_{l}({\bf k},t)-\int_{0}^{t}L_{KK}({\bf k}%
,t-t^{\prime })\chi _{KK}^{-1}\delta \sigma _{K}({\bf k},t)dt^{\prime }
\nonumber \\
&&-\int_{0}^{t}L_{UK}({\bf k},t-t^{\prime })\chi _{UU}^{-1}\delta \sigma
_{U}({\bf k},t)dt^{\prime }+f_{K}({\bf k},t) \label{dsigmakdt}
\end{eqnarray}
and
\begin{eqnarray}
\frac{\partial \delta \sigma _{U}({\bf k},t)}{\partial t} &=&ik\chi
_{UU}\chi _{jj}^{-1}\delta j_{l}({\bf k,}t{\bf )}-\int_{0}^{t}L_{UU}({\bf k}%
,t-t^{\prime })\chi _{UU}^{-1}\delta \sigma _{U}({\bf k},t)dt^{\prime }
\nonumber \\
&&-\int_{0}^{t}L_{UK}({\bf k},t-t^{\prime })\chi _{KK}^{-1}\delta \sigma
_{K}({\bf k},t)dt^{\prime }+f_{U}({\bf k},t). \label{dsigmaudt}
\end{eqnarray}
In these equations, only $L_{KK}({\bf k},t)$, $L_{UU}({\bf k},t)$, and $%
L_{UK}({\bf k},t)$ remain unknown.

The extended dynamic description provided by Eqs. (\ref{continuity}), (\ref{djdt1}), (\ref{dsigmakdt}),
and (\ref{dsigmaudt}) can now be contracted down to a single time-evolution equation involving only
$\delta n({\bf k},t)$ \cite{delrio}. This essentially amounts to formally eliminating the variables $\delta j({\bf k},t),\ \delta
\sigma _{K}({\bf k},t)$, and $\delta \sigma _{U}({\bf k},t)$, from this system of equations. The result of such contraction procedure reads \cite{scgle0}
\begin{equation}
\frac{\partial \delta n({\bf k},t)}{\partial t}=-\int_{0}^{t}L(k,t-t^{\prime
})\chi _{nn}^{-1}\delta n({\bf k},t^{\prime })dt^{\prime }+f({\bf k},t), \label{cont2}
\end{equation}
where $f({\bf k},t)$ is a random term with zero mean and time-dependent
correlation function $\left\langle f({\bf k},0)f(-{\bf k},0)\right\rangle
=L(k,t)$ with $L(k,t)$ given, in Laplace space, by
\begin{equation}
L(k,z)=\frac{k^{2}\chi _{jj}}{z+z^{(s)}+\chi _{jj}^{-1}\Delta L_{jj}(k,z)}.
\end{equation}
with $z^{(s)}\equiv \zeta^{(s)}/M$ and
\begin{equation}
\Delta L_{jj}(k,z)=\frac{k^{2}\chi _{KK}}{z+L_{KK}\chi _{KK}^{-1}}+\frac{%
k^{2}\chi _{UU}\left[ 1-\frac{L_{KU}\chi _{UU}^{-1}}{z+L_{KK}\chi _{KK}^{-1}}%
\right] ^{2}}{z+L_{UU}\chi _{UU}^{-1}-\frac{\chi _{KK}^{-1}L_{KU}L_{UK}\chi
_{UU}^{-1}}{z+L_{KK}\chi _{KK}^{-1}}}
\end{equation}

Multiplying Eq. (\ref{cont2}) by $\delta n(-{\bf k},0)$, and taking the equilibrium average, this equation becomes the time-evolution equation for the intermediate scattering function $F(k,t)\ \equiv\ \langle\delta n({\bf k},t)\delta n(-{\bf k},0)\rangle$, an equation that can be written as an expression for the Laplace transform   $F(k,z)$ in terms of the memory functions $L_{KK}(k,z)$, $L_{KU}(k,z)=L_{UK}(k,z)$, and $L_{UU}(k,z)$, namely,
\begin{equation}
F(k,z)=\frac{S(k)}{z+\frac{k^{2}S^{-1}(k)\chi _{jj}}{z+z^{(s)}+\frac{k^{2}\chi
_{jj}^{-1}\chi _{KK}}{z+L_{KK}\chi _{KK}^{-1}}+\frac{k^{2}\chi
_{jj}^{-1}\chi _{UU}\left[ 1-\frac{L_{KU}\chi _{UU}^{-1}}{z+L_{KK}\chi
_{KK}^{-1}}\right] ^{2}}{z+L_{UU}\chi _{UU}^{-1}-\frac{\chi
_{KK}^{-1}L_{KU}L_{UK}\chi _{UU}^{-1}}{z+L_{KK}\chi _{KK}^{-1}}}}}.
\end{equation}
At this point we can discuss the limit of vanishing solvent friction, $\zeta^{(s)} \to 0$. As discussed above, in this limit our Brownian fluid becomes a Newtonian system, in the sense that its microscopic dynamics is described by Eq. (\ref{eq1}) without the friction and fluctuating terms.  Thus, the expression for $F(k,z)$ describing the collective dynamics of an atomic liquid can be obtained from the previous expression by simply setting $z^{(s)}=0$, i.e,
\begin{equation}
F(k,z)=\frac{S(k)}{z+\frac{k^{2}S^{-1}(k)\chi _{jj}}{z+\frac{k^{2}\chi
_{jj}^{-1}\chi _{KK}}{z+L_{KK}\chi _{KK}^{-1}}+\frac{k^{2}\chi
_{jj}^{-1}\chi _{UU}\left[ 1-\frac{L_{KU}\chi _{UU}^{-1}}{z+L_{KK}\chi
_{KK}^{-1}}\right] ^{2}}{z+L_{UU}\chi _{UU}^{-1}-\frac{\chi
_{KK}^{-1}L_{KU}L_{UK}\chi _{UU}^{-1}}{z+L_{KK}\chi _{KK}^{-1}}}}}. \label{fdkz0}
\end{equation}

In a completely analogous manner we can derive the corresponding expression for
the \emph{self}-ISF $F_S(k,t)$, with the following result
\begin{equation}
F_S(k,z)=\frac{1}{z+\frac{k^{2}\chi _{jj}}{z+\frac{k^{2}\chi
_{jj}^{-1}\chi _{KK}}{z+L_{KK}^{(S)}\chi _{KK}^{-1}}+\frac{k^{2}\chi
_{jj}^{-1}\chi _{UU}^{(S)}\left[ 1-\frac{L_{KU}^{(S)}\chi _{UU}^{(S)-1}}{z+L_{KK}^{(S)}\chi
_{KK}^{-1}}\right] ^{2}}{z+L_{UU}^{(S)}\chi _{UU}^{(S)-1}-\frac{\chi
_{KK}^{-1}L_{KU}^{(S)}L_{UK}^{(S)}\chi _{UU}^{(S)-1}}{z+L_{KK}^{(S)}\chi _{KK}^{-1}}}}}, \label{fsdkz0}
\end{equation}
with
\begin{equation}
\chi^{(S)} _{UU}\equiv \frac{n\chi _{jj}^{2}}{k^{2}}\left[ \int
d{\bf r}g(r)\left(\frac{\partial ^{2}\beta u(r)}{\partial z^{2}}
\right) \right]. \label{chiuself}
\end{equation}
These general results now will serve as the basis for the analysis of the long-time dynamics of an atomic liquid, carried out in the following section.

\section{Long-time dynamic equivalence of atomic and colloidal liquids.}\label{sectionIII}

In this section we analyze the long-time (or small frequency) limit of the general expressions for $F(k,z)$ and $F_S(k,z)$ in Eqs. (\ref{fdkz0}) and (\ref{fsdkz0}). With this purpose,
as an additional approximation (following Ref. \cite{scgle0}, but introduced here only for simplicity) let us first neglect the possible crossed kinetic couplings represented by the memory
functions $L_{KU}(k,z)=L_{UK}(k,z)$ in this equation. This leads
to simpler expression for the ISF of an atomic
liquid, namely,
\begin{equation}
F(k,z)=\frac{S(k)}{z+\frac{k^{2}S^{-1}(k)\chi
_{jj}}{z+\frac{k^{2}\chi _{jj}^{-1}\chi _{KK}}{z+L_{KK}(k,z)\chi
_{KK}^{-1}}+\frac{k^{2}\chi _{jj}^{-1}\chi _{UU}}{z+L_{UU}(k,z)\chi
_{UU}^{-1}}}} \label{fkz}
\end{equation}
and
\begin{equation}
F_S(k,z)=\frac{1}{z+\frac{k^{2}\chi _{jj}}{z+\frac{k^{2}\chi
_{jj}^{-1}\chi _{KK}}{z+L_{KK}^{(S)}(k,z)\chi
_{KK}^{-1}}+\frac{k^{2}\chi _{jj}^{-1}\chi
_{UU}^{(S)}}{z+L_{UU}^{(S)}(k,z)\chi _{UU}^{(S)-1}}}}. \label{fskz}
\end{equation}

Eqs. (\ref{fkz}) and (\ref{fskz}) express $F(k,t)$ and  $F_S(k,t)$ in terms of the unknown memory
functions $L_{KK}(k,z)$, $L_{UU}(k,z)$, $L_{KK}^{(S)}(k,z)$ and
$L_{UU}^{(S)}(k,z)$. To understand the properties of these memory functions, with the aim of introducing additional approximations or simplifications, it helps to analyze their physical meaning. For this, let us recall that the memory functions
$L_{KK}(k,z)$ and $L_{KK}^{(S)}(k,z)$ are associated with the
relaxation of the kinetic part $\sigma_K ^{{\alpha \beta }}({\bf
k},t)\equiv N^{-1/2}\sum_{i=1}^{N} v_{i}^{\alpha }v_{i}^{\beta }
e^{i{\bf k\cdot r}_{i}(t)}$ of the stress tensor, whose trace
$\sigma_K ({\bf k},t)\equiv N^{-1/2}\sum_{i=1}^{N}
\textbf{v}_{i}^{2}  e^{i{\bf k\cdot r}_{i}(t)}$ is directly related
with the FT of the local kinetic energy density. Thus, $L_{KK}(k,z)$
and $L_{KK}^{(S)}(k,z)$ essentially describe the transport of
molecular kinetic energy, i.e., the transport of heat. These
transport processes occur primarily by means of molecular collisions
and quickly lead to a uniform distribution of the mean kinetic
energy of the particles, i.e., to thermal (but not thermodynamic!)
equilibrium. As a result, these memory functions may be expected to
be related with heat conductivity, and to decay within molecular
collision times. The memory functions $L_{UU}(k,z)$ and
$L_{UU}^{(S)}(k,z)$, on the other hand, describe the relaxation of
the configurational component of the stress tensor, which involves
structural relaxation processes that may decay after much longer
relaxation times.

Because of this, if one is interested in the long-time behavior of the ISFs,
one may neglect the frequency-dependence of
$L_{KK}(k,z)$, and replace it by its zero-frequency limit,
\begin{equation}
L_{KK}(k,z)\approx L_{K}(k) \equiv \lim _{z\to 0} L_{KK}(k,z)
\end{equation}
in Eq. (\ref{fkz}), and similarly for $L_{KK}^{(S)}(k,z)$,
\begin{equation}
L_{KK}^{(S)}(k,z)\approx L_{K}^{(S)}(k) \equiv \lim _{z\to 0} L_{KK}^{(S)}(k,z),
\end{equation}
in Eq. (\ref{fskz}). In addition, we also assume that the kinetic
coefficients $L_{K}(k)$ and $L_{K}^{(S)}(k)$ are not fundamentally
different from each other, so that we neglect their possible
differences,
\begin{equation}
L_{K}(k) \approx L_{K}^{(S)}(k).
\end{equation}
At this point we take the desired long-time limit $t>>\tau_0$ in
the resulting approximate expressions for $F(k,z)$ and $F_S(k,z)$.
This amounts to neglecting the frequency $z$ compared with the
frequencies $z_D \equiv L_{KK}^{(S)}(k,z)\chi_{KK}^{-1}$ and $z_B
\equiv k^{2}\chi _{jj}^{-1}\chi _{KK}/z_D$ in  Eqs. (\ref{fkz})
and (\ref{fskz}), which leads to the ``overdamped" form of these
expressions, namely,
\begin{equation}
F(k,z) = \frac{S(k)}{z+\frac{k^{2}S^{-1}(k)D^0}{1+C(k,z)}}
\label{fkz2}
\end{equation}
and
\begin{equation}
F_S(k,z) = \frac{1}{z+\frac{k^{2}D^0}{1+C_S(k,z)}}, \label{fskz2}
\end{equation}
where we have defined the memory functions $C(k,z)$ and $C_S(k,z)$
as
\begin{equation}
C(k,z) \equiv \left[\frac{k^{2}D^0\chi _{jj}^{-2}\chi
_{UU}}{z+L_{UU}(k,z)\chi _{UU}^{-1}}\right] \label{ckz}
\end{equation}
and
\begin{equation}
C_S(k,z) \equiv \left[\frac{k^{2}D^0\chi _{jj}^{-2}\chi
_{UU}^{(S)}}{z+L_{UU}^{(S)}(k,z)\chi _{UU}^{(S)-1}}\right],
\label{cskz}
\end{equation}
respectively.

In these equations we have denoted the unknown frequency $z_D=L_{K}^{(S)}(k)\chi _{KK}^{-1}$  as
\begin{equation}
L_{K}^{(S)}(k)\chi _{KK}^{-1}= 2k^2D^0. \label{dkinetictheory3}
\end{equation}
The use of the symbol $D^0$ is, of course, not accidental, since this parameter can be identified with the self-diffusion coefficient that describes the sequence of ballistic random flights of a tracer particle as it collides with its neighbor particles. To see this, notice that in the conditions in which the effects of the configurational memory function $C_S(k,z)$ are negligible (such as in the low-density regime, in which $\chi
_{UU}^{(S)}=\chi_{UU}=0$), Eq. (\ref{fskz2}) becomes
\begin{equation}
F_S(k,z) \approx \frac{1}{z+k^2D^0},
\label{fkz0}
\end{equation}
or
\begin{equation}
F_S(k,t) \approx e^{-k^{2}D^0 t}. \label{fkt00}
\end{equation}
This result implies that the MSD is given by $W(t)\approx D^0t$, i.e., that the motion of a tracer particle after many collision times will be diffusive. The corresponding diffusion coefficient $D^0$ must then be identical to that determined by kinetic-theoretical arguments, i.e., must be given by $D^0 = (l_0)^2/\tau_0$, where $l_0$ and $\tau_0$ are, respectively, the mean free path and the mean free time. Since $l_0/\tau_0 =v_0$, $D^0$ can also be written as $D^0 = v_0l_0$. If we then estimate the mean free path  $l_{0}$ to be given by $l_{0} \sim
1/n\sigma^2$, with $n\equiv N/V$ and with $\sigma$ being the
collision diameter of the particles, we then have that $D^0
\sim \sqrt{k_BT/M}/(n\sigma^2)$. In fact, the rigorous value of
$D^0$ is  \cite{chapmancowling}
\begin{equation}
D^0\equiv \frac{3}{8\sqrt
\pi}\left(\frac{k_BT}{M}\right)^{1/2}\frac{1}{n\sigma^2}.
\label{dkinetictheory}
\end{equation}

The comparison of the overdamped expressions for  $F(k,z)$ and $F_S(k,z)$ in Eqs. (\ref{fkz2})-(\ref{cskz}) above, with the corresponding overdamped  results of a colloidal liquid (i.e., with Eqs. (4.24) and (4.33) of Ref. \cite{scgle0}), reveals the
remarkable formal identity between the long-time expressions for $F(k,t)$ and $F_S(k,t)$ of an atomic liquid, and the corresponding results for the analogous colloidal system. The fundamental difference between these two cases is to be found in the definition of the diffusion coefficient $D^0$, which in the present (atomic) case depends on temperature and density, and is given by the kinetic-theoretical result in Eq. (\ref{dkinetictheory}), whereas in colloidal liquids it is a constant, identical to the short-time self-diffusion coefficient given, for example, by the Einstein-Stokes expression in the absence of hydrodynamic interactions. Thus, this formal identity implies that
the long-time dynamic properties of an atomic liquid will then
coincide with the corresponding properties of a colloidal system
with the same $S(k)$, provided that the time is scaled as $D^0t$, with the respective meaning and definition of $D^0$. This observation has important implications, which can be tested, for example, by comparing the simulation results for $F_S(k,t)$ obtained by both, molecular dynamics and Brownian dynamics, for the same system and conditions.

\section{Test of the predicted long-time dynamic equivalence.}\label{sectionIV}

In this section we perform the test of the predicted long-time dynamic equivalence between a model atomic liquid and its corresponding Brownian fluid. This dynamic equivalence is tested here by comparing the macroscopic dynamics of the hard sphere liquid when the motion of its constituent particles is described, respectively, by Eqs. (\ref{eq1}) without and with the solvent friction terms present, i.e., by performing and comparing the molecular and the Brownian dynamics simulations of these properties.

\begin{figure}[ht]
\begin{center}
\includegraphics[scale=.35]{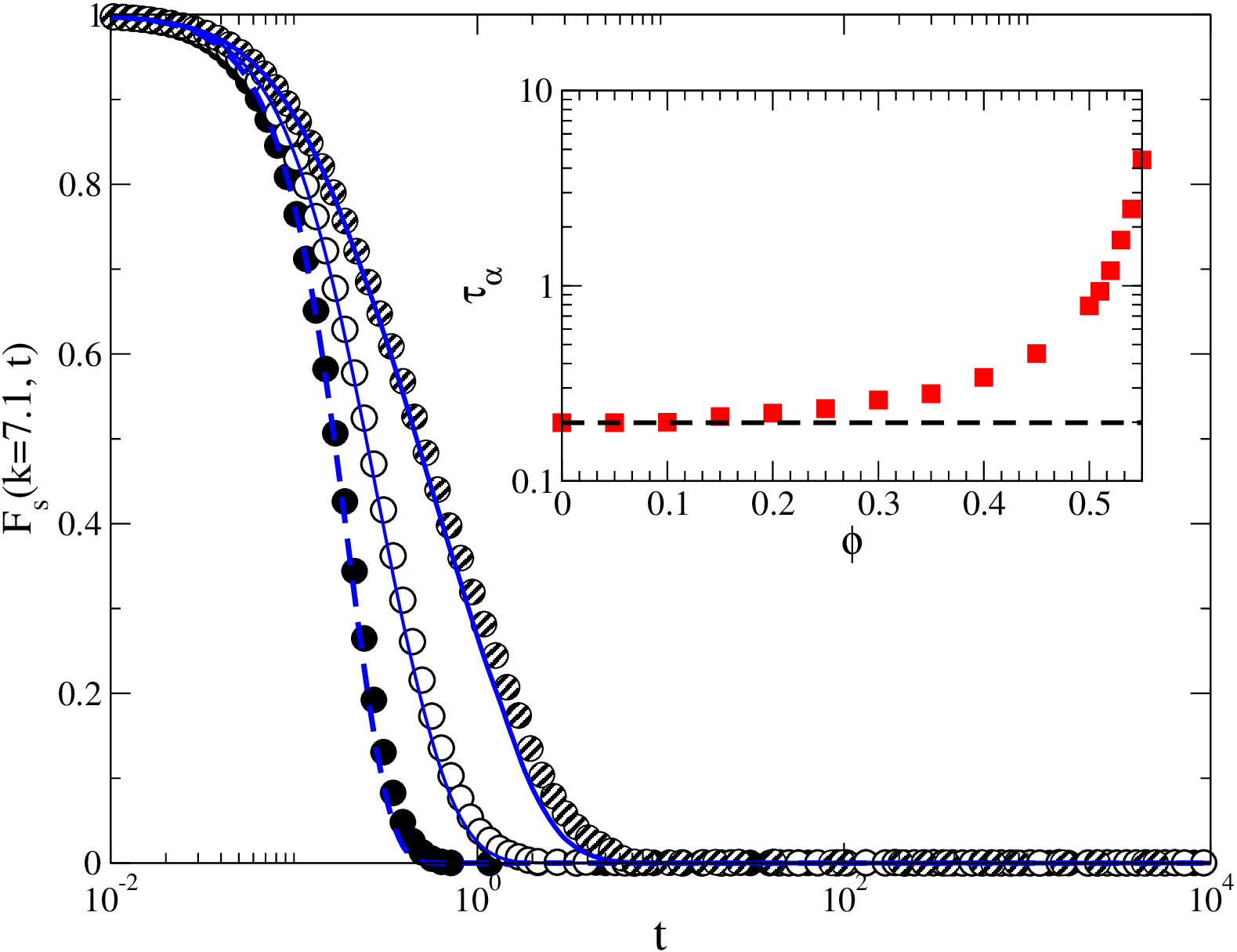}
\caption{Molecular dynamics results for the self-ISF $F_S(k,t)$ of a
hard-sphere fluid as a function of time $t$ (expressed in ``molecular" units $[\sigma/v_0]$) at fixed wave-vector $k\sigma=7.1$, and volume fractions $\phi$ = 0.1 (solid circles), 0.4 (empty circles), and 0.5 (striped circles). The dashed line is the exact limit $F_S(k,t)=\exp(-\frac{1}{2}k^{2}v_{0}^{2}t^{2})$ and the solid lines are the results of the Gaussian approximation $F_S(k,t) \approx \exp[-k^2 W(t)]$, with $W(t)$ given by the same molecular dynamics simulation data. In the inset we plot the relaxation time $\tau_{\alpha}$ (also in units of $[\sigma/v_0]$), defined by the condition  $F_{S}(k,\tau_{\alpha})=1/e$, for these and other volume fractions; the horizontal dashed line indicates the limiting value $v_{0}\tau_{\alpha}/\sigma=\frac{\sqrt{2}}{k\sigma}=0.199$.  } \label{fig1}
\end{center}
\end{figure}

As a reference let us first recall the exact short-time limit of the self-ISF of  an atomic liquid.  Since for correlation times $t$ shorter than the mean free time $\tau_0$ all the particles move ballistically,  $[{\bf r}_i(t)-{\bf r}_i(0)]={\bf v}_i^{0}t$, we have that $F_S(k,t)\approx (1/N)\langle\sum_{i=1}^N \exp {[i{\bf k}\cdot{\bf v}_i(0)t]}\rangle$. Using the equilibrium distribution of the initial velocities ${\bf v}_i(0)$, one can see that the exact short-time limit of the self-ISF is given by  $F_S(k,t)=\exp(-\frac{1}{2}k^{2}v_{0}^{2}t^{2})$. This expression provides an excellent approximation at small volume fractions, where $F_S(k,t)$ has decayed to negligible values for $t\approx \tau^0$, as illustrated by its comparison in the main panel  of Fig. (\ref{fig1}) with the MD-simulated $F_S(k,t)$ for the hard sphere fluid at $\phi =0.1$ and $k\sigma=7.1$.

The MD simulations were conducted on a soft-sphere system, and the results were then mapped onto those of the equivalent hard-sphere liquid as discussed in Ref. \cite{soft2}. The soft-sphere simulations were carried out using the velocity-verlet algorithm  with $N=1000$ particles of the same mass $M$ in a volume $V$ and a
time step $\Delta t/t^{*}=1 X 10^{-3}\sqrt{m\sigma^{2}/\epsilon}$.  During the
equilibration and production cycles, temperature was kept
constant by a simple rescaling of the velocities of the particles every 100
time steps. For high volume fractions we used polydisperse systems, where the diameters of the $N$ particles were
evenly distributed between $\overline{ \sigma} (1-w/2)$ and
$\overline{ \sigma} (1+w/2)$, with $\overline \sigma$ being the mean
diameter. We consider the case $w=0.3$, corresponding to a
polydispersity $s_\sigma = w/\sqrt{12}=0.0866$. The length, mass, and time
units employed are, respectively, $\overline{\sigma}$, $M$, and
$\overline{\sigma}\sqrt{M/k_BT}$. The simulations are carried out for an array of volume fractions $\phi=(\pi/6) n \overline{\sigma^3}$ where
$\overline{\sigma^3}$ is the third moment of the size distribution and $n$ is the total number density $n\equiv N/V$.

Defining the relaxation time $\tau_{\alpha}$ by the condition  $F_{S}(k,\tau_{\alpha})=1/e$, we have that in the ballistic regime $\tau_{\alpha}$ can be approximated by $\tau_{\alpha}=\frac{\sqrt{2}}{kv_{0}}$, which is the low-density limiting value represented in the inset of Fig.  \ref{fig1} by the horizontal dashed line. The inset also plots the simulation results for $\tau_{\alpha}$ in a wide range of volume fractions, to show the deviations from this limiting behavior as the density is increased. Beyond this low-density regime, these deviations become increasingly more important, as also illustrated in the main panel of Fig.  \ref{fig1} by the MD simulation results for $F_S(k,t)$ at the near-freezing volume fractions $\phi=0.4$ and 0.5. Here, of course, the $\phi$-independent limit  $F_S(k,t)=\exp(-\frac{1}{2}k^{2}v_{0}^{2}t^{2})$ is clearly inadequate, although the Gaussian approximation, $F_S(k,t) \approx \exp[-k^2 W(t)]$ still provides an accurate representation of the  short-time decay of this function. This is illustrated by the solid lines of the main panel  of Fig. \ref{fig1}, which result from employing the MD-simulated data for the mean squared displacement $W(t)$  in $F_S(k,t)= \exp[-k^2 W(t)]$.

\begin{figure}
\begin{center}
\includegraphics[scale=.27]{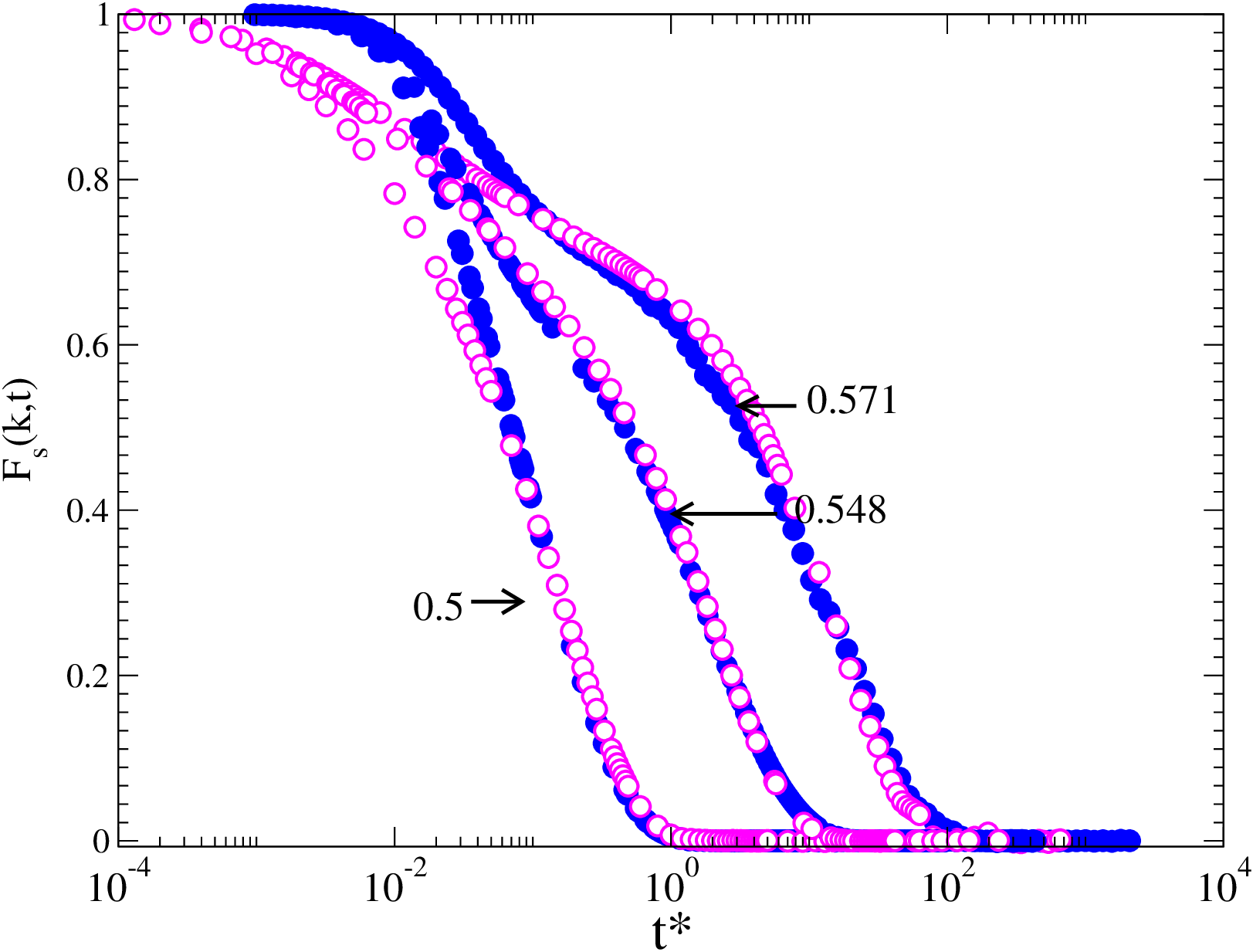}
\includegraphics[scale=.27]{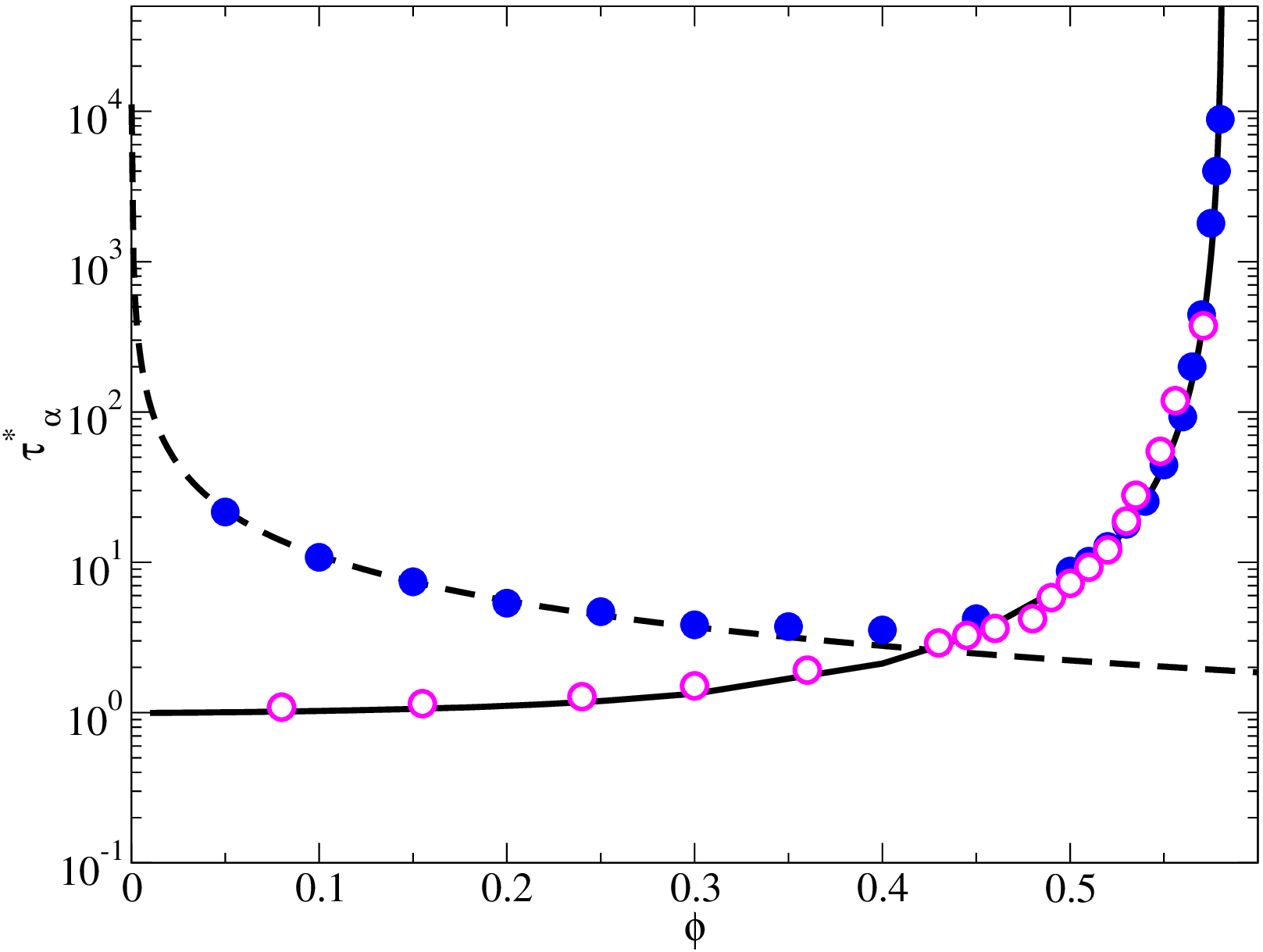}
\caption{(a) Molecular dynamics (solid symbols) and Brownian dynamics (empty symbols) simulation results for the self-intermediate scattering function $F_S(k,t)$ of the hard sphere liquid at  volume fraction $\phi=0.50$, 0.548, and 0.571, evaluated at the main peak of the static structure factor and plotted as a function of the dimensionless time  $t^*\equiv D^0t/\sigma^2$. (b) Volume fraction dependence of the dimensionless $\alpha$-relaxation time $\tau^*\equiv k^2D^0\tau_\alpha$ of the hard sphere liquid determined from the corresponding molecular dynamics (solid symbols) and Brownian dynamics (empty symbols) simulations. The dashed curve represents the low-density limit $\tau^* = (k\sigma)\sqrt{2 \pi}/16\phi$, whereas the solid curve correspond to the results of the SCGLE theory (Eqs. (1),(2),(5)-(8) of Ref. \cite{todos2}, with $k_c =
1.305(2\pi/\sigma))$.}
\label{fig2}
\end{center}
\end{figure}

With this low-density short-time ballistic limiting behavior as a reference, let us now  compare the simulation results for $F_S(k,t)$ obtained by both, molecular dynamics and Brownian dynamics simulations, for the same system and conditions. For this comparison, in addition to the molecular dynamics simulations, we performed Brownian dynamics simulations of the hard sphere liquid using the conventional Ermak and McCammon's Brownian dynamics algorithm \cite{tildesley,ermakmccammon} on a soft sphere fluid, and then mapping the results onto those of the hard-sphere liquid  according to the methodology proposed and explained in Ref. \cite{dynamicequivalence}. The resulting comparison provides a test of the theoretical prediction of the previous sections, that the dynamics of an atomic liquid coincides with the dynamics of the corresponding Brownian fluid in the \emph{opposite} regime, i.e., for high densities and long times. Thus,  Fig. \ref{fig2}(a), presents both simulation results for the hard sphere system at three volume fractions, $\phi=0.50$, 0.548, and 0.571, representing the metastable regime of the hard sphere liquid. As this figure illustrates, plotting  $F_S(k,t)$ as a function of the scaled time  $t^*\equiv D^0t/\sigma^2$ clearly exhibits the expected long-time  dynamic equivalence between atomic and Brownian liquids. We notice, however, that  this long-time dynamic equivalence is not observed in $F_S(k,t)$ at lower volume fractions, corresponding to the stable fluid regime ($\phi \lesssim 0.45$). The reason for this is that in such regime, illustrated in Fig. \ref{fig1}, the decay of $F_S(k,t)$ to a value $\approx e^{-1}$ occurs within times comparable to the mean free time $\tau_0$ and is, hence, intrinsically ballistic.  It is only at higher volume fractions that this long-time dynamic equivalence is fully exhibited by the \emph{diffusive} decay of $F_S(k,t)$, as illustrated by Fig. \ref{fig2}(a).

Another manner to summarize this observation is to compare the volume fraction dependence of the relaxation time $\tau_\alpha$ of both, molecular and Brownian dynamics. In Fig. \ref{fig2}(b) these simulation results are presented in terms of the dimensionless $\alpha$-relaxation time $\tau^*\equiv k^2D^0\tau_\alpha$. For a Brownian liquid $\tau_\alpha\to 1/ k^2D^0$ as  $\phi \to 0$, with a $\phi$-independent short-time diffusion coefficient $D^0$, so that  $\tau^*\to 1$ as $\phi \to 0$. As discussed in the previous section, however, for atomic liquids $\tau_{\alpha} \to \sqrt{2}/kv_{0}$  as $\phi \to 0$, so that in the same limit $\tau^* \to (k\sigma)\sqrt{2 \pi}/16\phi$, where we have taken into account the fact that in this case, the short-time diffusion coefficient $D^0$ is given by the kinetic-theoretical result in Eq. (\ref{dkinetictheory}). This  limiting behavior was represented  by  the horizontal dashed line of Fig. \ref{fig1}, and is now represented by the dashed curve of Fig. \ref{fig2}(b). From the comparison in this figure one can see that the long-time dynamic equivalence manifests itself in the collapse of the molecular and Brownian dynamics data for $\tau^*$ at high volume fractions. For smaller volume fractions, the differences in the short-time behavior of $F_S(k,t)$ lead to the observed differences between the molecular and Brownian dynamics results for $\tau^*$ below a crossover volume fraction  located near the freezing transition of the HS liquid.

The solid curve in Fig. \ref{fig2}(b) is the prediction for $\tau^*\equiv k^2D^0\tau_\alpha$ of the self-consistent generalized Langevin equation (SCGLE) theory of colloid dynamics, i.e., of Eqs. (1),(2),(5)-(8) of Ref. \cite{todos2}. These are actually Eqs. (\ref{fkz2}) and (\ref{fskz2}) above, complemented by the closure relation $C(k,t)=C_S(k,t)=\lambda(k)\Delta \zeta(t)$, where $\Delta \zeta(t)$ is the time-dependent friction function describing the configurational contribution to the friction force on a tracer particle (given by Eq. (6) of Ref. \cite{todos2}). The static structure factor of the hard sphere system, needed as an input in these equations, is provided by the Percus-Yevick approximation with its Verlet-Weis correction \cite{percusyevick,verletweis}.
The function $\lambda (k) = 1/[1+(k/k_c)^2]$ is a phenomenological ``interpolating" function, with the cutoff wave-vector $k_c$  used here to calibrate the SCGLE theory by optimizing the overall agreement of its predictions with the  data for $\tau^*$ constituted by the totality of the Brownian dynamics results (squares) and by the molecular dynamics data corresponding to the metastable liquid ($0.5\lesssim \phi$) in this figure. This  calibration procedure results in the value $k_c = 1.305(2\pi/\sigma)$.

As said above,  the short-time differences between the molecular and the Brownian dynamics data for $\tau^*$ in Fig. \ref{fig2}(b) appear at densities below a crossover volume fraction  located, for the data in this figure,  near the freezing transition of the HS liquid. The location of this crossover depends, however, on the wave-vector $k$ at which the decay of $F_S(k,t)$ is being observed, moving to a vanishing value in the long-wavelength limit, $k\to 0$. This means that in this limit the molecular and Brownian dynamics results for $\tau^*$ will be identical at all volume fractions. In fact, this is also what happens to the most representative long-time dynamic property, namely, the long-time self-diffusion coefficient. $D_L$ is defined as  $D_L \ \equiv \lim_{t \to \infty}
\langle(\Delta \textbf{r}(t))^2\rangle / 6t $, but is also given by $D_L= \lim_{k\to 0} \lim_{z\to 0} [k^2F_S(k,z)]^{-1}= D^0/[1+C_S(k=0,z=0)]$. According to Eq. (\ref{fskz2}) above, and within the SCGLE closure $C_S(k,t)=\lambda(k)\Delta \zeta(t)$, for an atomic system this parameter, scaled as $D^*\equiv D_L/D^0 $,  can be written as
\begin{equation}
D^*= 1/[1+\int _0^{\infty}\Delta \zeta^*(t)dt],
\label{dstar}
\end{equation}
with $\Delta \zeta^*(t)$ given, according to Eq. (6) of Ref. \cite{todos2}, by
\begin{equation}
\Delta \zeta ^*(t) =\frac{D^0}{3\left( 2\pi \right) ^{3}n}\int d
{\bf k}\left[\frac{ k[S(k)-1]}{S(k)}\right] ^{2}F(k,t)F_S(k,t).
\label{dzdt0p}
\end{equation}
These equations, however, are identical to their colloidal counterpart. Thus, they imply that the parameter $D^*$ of an atomic liquid  must be indistinguishable from the corresponding parameter of the equivalent colloidal system with the same interactions and the same static structure factor.

\begin{figure}
\begin{center}
\includegraphics[scale=.32]{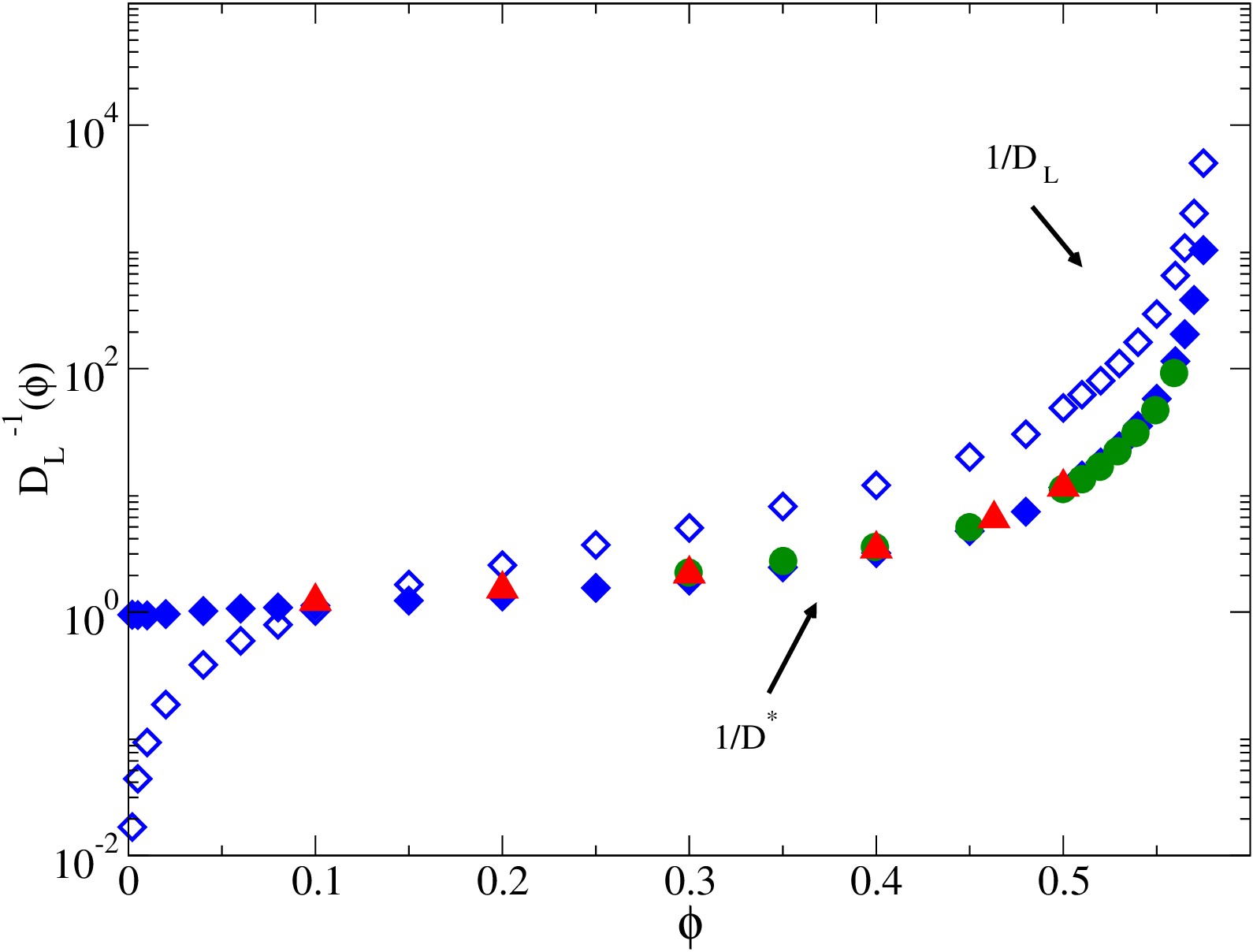}
\caption{ Long-time self-diffusion coefficient $D_L (\phi)$ of the
hard-sphere fluid determined by molecular
dynamics simulations \cite{gabriel0,gabriel}, expressed in ``atomic units" $\sigma(k_BT/M)^{1/2}$ (empty
diamonds), and normalized as  $D^* (\phi) \equiv D_L(\phi)/ D^0(\phi)$, with
$D^0(\phi)$ given by Eq. (\ref{dkinetictheory}) (full diamonds). The other full symbols
are the Brownian dynamics simulation results for $D^*$ from Refs.
\cite{cichocki} (triangles) and \cite{tokuyamabd} (circles).  }
\label{fig3}
\end{center}
\end{figure}

The accuracy of this important and distinct prediction can also be checked by comparing the corresponding molecular and Brownian dynamics results.
Thus, in Fig. \ref{fig3} we plot molecular dynamics data for
$D_L(\phi)$ of a hard-sphere fluid both, in the ``usual" atomic
units $ \sigma(k_BT/M)^{1/2}$, and scaled as $D^* (\phi) \equiv
D_L(\phi)/ D^0(\phi)$, with $D^0(\phi)$ given by Eq.
(\ref{dkinetictheory}). The same figure also presents available
Brownian dynamics simulation results for $D_L(\phi)$ of the hard
sphere system without hydrodynamic interactions, also scaled as $D^*
(\phi)\equiv D_L(\phi)/ D^0$, but with $D^0$ being the
$\phi$-independent short-time self-diffusion coefficient of the
Brownian particles. Clearly, the ``colloidal" and the ``atomic"
results for $D^*$ collapse onto the same curve, which we denote by
$D^*_{HS} (\phi)$. One immediate and important consequence of this comparison is, for example, that  L\"owen's dynamic criterion for freezing \cite{lowen} now applies for both, the atomic and the colloidal hard sphere liquid, i.e., the condition $D^*_{HS} (\phi)\approx 0.1$ occurs at $\phi=\phi_{HS}^{(f)}=0.494$ in both cases. The comparison in this figure, however, is only one particular manifestation of the more general long-time dynamic scaling suggested by the present work, whose applications were also illustrated by the other results presented in this section.

\section{Summary and discussion.}\label{sectionV}

In this paper we have discussed the relationship between the dynamics of atomic and Brownian liquids, by describing the macroscopic dynamics of both kinds of systems within the same theoretical formalism. We have based this discussion on the application of the generalized Langevin equation formalism to the derivation of general memory-function expressions for the (collective and self) intermediate scattering functions of an atomic liquid. The actual derivation, however, consisted in the review of the previous derivation \cite{scgle0} of the time-evolution equations for $F(k,t)$ and $F_S(k,t)$ of the corresponding Brownian fluid, keeping track of the effects of the solvent friction. At the end of such derivation  the zero-friction limit was taken, to obtain the corresponding  time-evolution equations for $F(k,t)$ and $F_S(k,t)$ of our atomic liquid (Eqs. (\ref{fdkz0}) and (\ref{fsdkz0})).

We then analyzed the long-time limit of these results for $F(k,t)$ and $F_S(k,t)$. The comparison of such overdamped expressions with the corresponding  results in the case of a colloidal liquid, revealed the
remarkable formal identity between the long-time expressions for $F(k,t)$ and $F_S(k,t)$ of  atomic and  colloidal liquids. As discussed in Sect. III, the fundamental difference between these two cases lies in the definition of the diffusion coefficient $D^0$; in atomic liquids it depends on temperature and density, and is given by the kinetic-theoretical result in Eq. (\ref{dkinetictheory}), whereas in colloidal liquids it is a constant, given by the density-independent Einstein-Stokes value in the absence of hydrodynamic interactions. Let us mention that this dynamic equivalence can also be inferred by the derivation of the (generalized) Langevin equation that describes the motion of representative tagged particles in an atomic liquid \cite{atomictracerdiff}. The atomic-to-Brownian long-time dynamic equivalence thus seems to be a very robust prediction, with important physical consequences. It implies, for example, that in an atomic system, the self-diffusion coefficient $D^{0}$ determined by kinetic theory  plays the same role as the short-time self-diffusion coefficient $D^{(s)}$ in colloidal liquids. It also implies that the long-time dynamic properties of an atomic liquid will
coincide with the corresponding properties of a colloidal system
with the same $S(k)$, provided that the time is scaled as $D^0t$, with the respective meaning and definition of $D^0$. 

In section IV we tested this observation by comparing the simulation results for $F_S(k,t)$ obtained by both, molecular dynamics and Brownian dynamics, for the hard sphere system. As mentioned at the end of the previous section, one important consequence is that  L\"owen's dynamic criterion for freezing \cite{lowen} now applies for both, the atomic and the colloidal hard sphere liquid. This result, taken together with the dynamic equivalence between soft- and hard-sphere liquids recently discussed in Ref. \cite{soft2}, further extends the application of this criterion to soft-sphere molecular liquids. The most relevant implications of this dynamic equivalence have been corroborated by the systematic comparisons between molecular and Brownian dynamics simulations of the sort illustrated in this paper. A summary of this analysis has been advanced in a recent brief communication \cite{atombrownequivletter}.

We should mention, in addition, that in reality the validity of the present dynamic correspondence between atomic and colloidal liquids should extend over to colloidal systems involving hydrodynamic interactions, provided that the corresponding effects enter only through the value of the short-time self-diffusion coefficient  $D^{(s)}(\phi)$, which should then play the role of a density-dependent $D^0$, as suggested in \cite{prlhi}.
Besides analyzing further these important predictions,  we are in the process of applying  the general expressions for $F(k,t)$ and $F_S(k,t)$ for an atomic liquid derived in this paper, to the development of a self-consistent scheme to calculate these properties. The intention is to extend to atomic liquids the  self-consistent generalized Langevin equation (SCGLE) theory of colloid dynamics \cite{scgle1,scgle2}, including the description of dynamic arrest phenomena \cite{rmf,todos1,todos2} and the recently developed first-principles theory of equilibration and aging  \cite{noneqscgle0,noneqscgle1}. This, however, will be reported separately.

\bigskip

ACKNOWLEDGMENTS: The authors are grateful to G. P\'erez-\'Angel for providing the molecular dynamics data in Fig. 3, and to L. E.  S\'anchez-D\'iaz, P. Mendoza-M\'endez, and A. Vizcarra-Rend\'on, for valuable discussions. L. L.-F. and M. M.-N. acknowledge the kind hospitality of the Joint Institute for Neutron Sciences (Oak Ridge, TN), where part of this manuscript was written. We are grateful to W.-R. Chen and T. Egami for stimulating discussions.
This work was supported by the Consejo Nacional de
Ciencia y Tecnolog\'{\i}a (CONACYT, M\'{e}xico) through grants 84076 and
132540 and through the Red Tem\'atica de la Materia Condensada Blanda.

\vskip.5cm

%\begin{references}

\end{document}